# Age of Big Data and Smart Cities: Privacy Trade-Off

Kaoutar Ben Ahmed[#1], Mohammed Bouhorma[#2], Mohamed Ben Ahmed[#3]

[1]*Ph.D. Student,*[2]*Professor,*[3]*Assistant Professor,*

[#]*Laboratory of Informatics, Systems and Telecommunications*

*FSTT Faculty, Abdelmalek Essaadi University, Morocco*

*Abstract*— Data will soon become one of the most precious treasures we have ever had, 43 TRILLION GIGABYTES of data will be created by 2020 according to a study made by Mckinsey Global Institute, it's estimated that 2.3 TRILLION GIGABYTES of data is created each day and most companies in the US have 100.000 GIGABYTES of data stored. Data is recorded, stored and analyzed to enable technology and services that the world relies on every day, this technology is getting smarter and we will be soon living in a world of smart services or what is called smart cities. This article presents an overview of the topic pointing to its actual status and forecasting the crucial roles it will play in the future, we will define big data analytics and smart cities and talk about their potential contributions in changing our way of living and finally we will discuss the possible down side of this upcoming technologies and how it can fool us, violate our privacy and turn us into puppets or technology slaves.

## I.  INTRODUCTION

According to statistics from International Telecommunication Union (ITU) for July 2014, around 40% of the world population has an internet connection today with a number of approximately 2,949,533,612 Internet users in the world. In 1995, the percentage was less than 1%.

In one second 7,474Tweets sent, 1,229Instagram photos uploaded, 1,430Tumblr posts, 1,508Skype calls, 22,810GB of Internet traffic, 2,325,044Emails sent![1] These aren't just numbers but its years from our lives that we spend online exchanging personal information.

Actually we are living in an age of technology revolution, the services provided to users are getting smarter, not only smart technology age but smart transportation, smart agriculture, smart energy, smart education, smart government, smart homes…it's the age of smart cities.

So what is big data? How will the smart cities look like? How are we, as end users, interacting with this new era so far? Are we smart enough to handle smart services? Will we be in control of our personal lives or are we going to be controlled? These are questions we will be discussing through this state of the art article.

## II.  BIG DATA

A. *definition*

Big data is a term with no set definition, mainly because the meaning of "big" changes with the advance of technology. A decade ago, big data was measured in terabytes (or 1,000 to the fourth power in the International System of Units), and today the measure has reached petabytes, or 1,000 times that size. Soon, big data will likely mean Exabytes—or 1 million terabytes.[2] The term "big data" first emerged in the 1980s to describe the impact computers had on the social sciences in the 1960s and 1970s.[3]

Doug Laney, an analyst with the Meta Group, publishes in 2001 a research note titled "3D Data Management: Controlling Data Volume, Velocity, and Variety." Where he presents 3Vs that have become the defining three dimensions of big data: Volume, Velocity and Variety [4]:

Volume: The increasing volume and detail of information captured by enterprises, the rise of multimedia, social media, and the Internet of Things will fuel exponential growth in data for the foreseeable future. [5] Even modern machines such as cars, trains, power stations and planes all have increasing numbers of sensors constantly collecting masses of data. It is common to talk of having thousands or even hundreds of thousands of sensors all collecting information about the performance and activities of a machine. [6]

Variety: data sources are becoming diverse and new data types have appeared, we are not talking anymore about the traditional structured data that organizations used to deal with such as financial transactions, stock records, personal files and relational databases with well-defined field' types but today's sources are audio, video, photos, location data, Tweets, sensors' data and other unstructured and complex data types that are pushing to the limit the common and traditional computing technologies to process it.

Velocity: is the rate at which data is generated and changed, The New York Stock Exchange captures 1TB of trade information during each trading session, Experiments at the Large Hadron Collider at CERN, Europe's particle-physics laboratory near Geneva, generate 40 terabytes every second[7]. In addition, users increasingly want streaming data to be delivered to them in real time, and often on mobile devices. Online video, location tracking, augmented reality and many other applications now rely on large quantities of such high velocity data streams [8].

In addition to the three V's, some add a fourth to the big data definition: Veracity is an indication of data integrity and the ability for an organization to trust the data and be able to confidently use it to make crucial decisions. [9]

B. *Data Analytics*

Data analytics is the science of collecting, storing, extracting, cleansing, transforming, aggregating and analyzing data with the purpose of discovering and communicating meaningful information, thus converting that analysis into value-creating action. Data analytics is used in fields as varied as science and sports, advertising and public health. "It's a revolution," says Gary King, director of Harvard's Institute for Quantitative Social Science. "We're really just getting under way. But the march of quantification, made possible by enormous new sources of data, will sweep through academia, business and government. There is no area that is going to be untouched."[10]

Analytics have been used in business since the time management exercises that were initiated by Frederick Winslow Taylor in the late 19th century. Henry Ford measured pacing of assembly line. But analytics began to command more attention in the late 1960s when computers were used in decision support systems. Since then, analytics have evolved with the development of enterprise resource planning (ERP) systems, data warehouses, and a wide variety of other hardware and software tools and applications.[11] Most companies have realized lately the inevitable need to data and analytics to improve productivity and competitiveness. "Analytics will define the difference between the losers and winners going forward," says Tim McGuire, a McKinsey director. [12] Analytics uses descriptive and predictive models to gain valuable knowledge from data and uses this insight to recommend action or to guide decision making as well as optimizing and modelling the future.

Big data analytics applications follows the life cycle presented in figure 1, data can have a variety of sources which can be traditional and structured sources (RDBMS, OLAP cubes…) or unstructured sources (Web logs, Emails, Sensors, Social media…). Data must be optimized using a process of ingestion, manipulation, integration, cleansing and transformation of data into an optimal format for analysis, then analytics techniques are applied to describe the past and present situation and to predict and model the future.

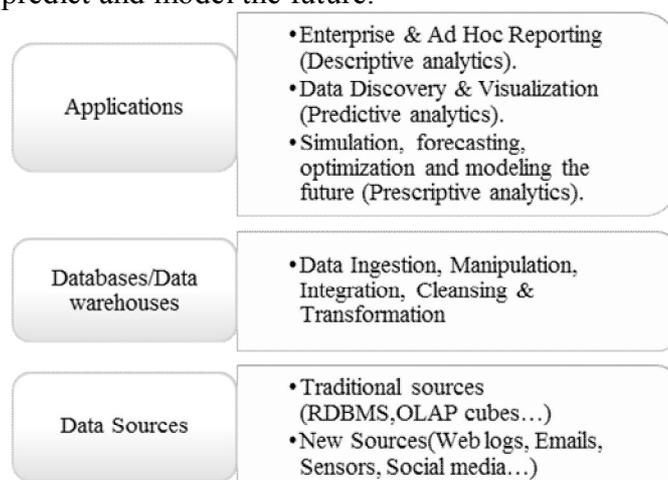

Fig1: Big data analytics life cycle

### III. SMART CITIES

Advances in Information and Communication Technologies (ICTs) are triggering a transformation of the environments where we live into intelligent entities globally known as Smart Spaces (Smart Homes, Smart Buildings, Smart Cities, etc.). They capture information using large sensors networks distributed throughout its domain (a house, a building, a whole city, etc.) and use it to intelligently adapt their behavior to the needs of the users.[13] It broadly refers to a city that is using new (ICTs) innovatively and strategically to achieve its aims.[14]

Giffinger et al. [15] has identified six dimensions of a smart city: smart economy; smart mobility; smart environment; smart people; smart living; and, finally, smart governance.

A city is considered to be smart when investments in human and social capital and traditional (transport) and modern (ICT) communication infrastructure fuel sustainable economic growth and a high quality of life, with a wise management of natural resources, through participatory governance. [16]

There are plenty of examples already out there of cities using big data to become more efficient. Philadelphia, for instance, estimates that it is saving $1 million (around £600,000) every year from fitting rubbish bins with sensors that indicate when the bin is full, thereby reducing the number of collections required.

In London, 500 datasets have been made available on a public website via a city dashboard that shows a range of data such as air pollution, crime statistics and even enables members of the public to track the real-time location of buses.

Furthermore, Glasgow is currently piloting a project where street lights have been programmed to increase in brightness if the noise level in the area rises. The lights are also integrated with monitored CCTV cameras so staff will quickly be able to spot if the rise in noise level is due to a problem or disturbance. [17]

Data is communicated to IT centers for analytics and information provisioning to citizens and other agencies. A strong ICT system is fundamental to the smart city as huge amounts of data need to be transported, stored and analyzed to extract information and act on it. It would indeed be necessary to set up a software platform to which data from disparate systems can be seamlessly transferred, quickly analyzed and actions taken.[18]

### IV. DISCUSSION: SMART SERVICES FOR SMART USERS: WILL WE BE IN CONTROL OF OUR CHOICES OR BE CONTROLLED?

After introducing these two concepts – big data and smart cities - and presenting the huge new possibilities that they are going to create in our lives, it's time to take a look into the future but this time let's do it from another angle. We will try in this chapter to put the puzzle pieces together so that we can see the whole image as clear as possible. What price are we going to pay in order to get these benefits? And is it really worthy? Does the rise of big data mean the downfall of privacy? We will be covering the big data collected about individuals, how it could be used in some future "smart" services that can turn us into a population "stupid" enough to be mind-controlled.

Individuals are identified by various ways namely biographic information, biometric data (face, fingerprint, Iris scans…), behavior data, travel data and banking information. Governments have always been obsessed with collecting data about citizens, for instance the East German state security agency known as Stasi until the fall of the Berlin Wall in 1989 masterminded a system of surveillance and spying on its citizens, they used both technical instruments and Human intelligence, telephones were wiretapped, special machines for letters opening, taking people's pictures and even their smell! Spies and agents were reporting not only what people are doing but also what are they thinking and planning. [19]

At that time, the web and smartphones were not yet invented but today technology made it even easier to collect, store and more importantly to analyze these data. The Executive Office of the US President BARACK OBAMA explains in its report about big data some samples of the data that the government is collecting about each individual: "The fields in each database are grouped into three categories: core biographical data, such as name, date of birth, and citizenship status; extended biographical data, including addresses, phone number, and email; and detailed encounter data derived from electronic and in-person interactions with the Department of Homeland Security. Encounter data is the most sensitive category. It may contain a law enforcement officer's observations about an individual they interview as well as allegations of a risk to homeland security they may pose." [20]

Not only governments that are interested in gathering individuals' data but also commercial firms, IT corporations, even a small amateur website enjoys storing user's web data. In the age of big data and smart services, the tracking of our web browsing, our social medial activities, records of our purchases, logs of our geographic locations transmitted by our smartphones and government snoops can reveal information about us more than what we could imagine.

Something as simple as scanning for wireless networks that we do every day with our mobile phones without knowing the fact that we are also beaming out a list of networks we've previously connected to, even when we're not using wireless actively, this data can easily reveal information about places we have visited, work locations, hotels we had been to recently, home addresses, even our names sometimes.[21]

In July 1993, The New Yorker published a cartoon by Peter Steiner that depicted a Labrador retriever sitting on a chair in front of a computer, paw on the keyboard, as he turns to his Beagle companion and says, "On the Internet, nobody knows you're a dog." Two decades later, interested parties not only know you're a dog, they also have a pretty good idea of the color of your fur, how often you visit the vet, and what your favorite doggy treat is. Web browsing is far away from preserving user's privacy, instead a big industry is operating in the background called online tracking, the companies specialized in this activity had developed a robust and practically undefeatable techniques to complete their mission which is identifying who is sitting behind the computer screen, what is his interests and activities. Device fingerprinting is a form of advanced online tracking that enables companies to spy on people even when they configure their browsers to avoid being tracked and it collects and identifies information about unique characteristics of the individual computers people use. Under the assumption that each user operates his or her own hardware, identifying a device is tantamount to identifying the person behind it. [22]

Mobile apps that we do trust and install have access to the majority – if not all - of our personal information such as Contacts, calendar, SMS, call log, media files, phone number, device IDs, location and have control on functionalities inside our devices like Camera and Microphone. Users give voluntary these apps the access permissions and most of them do not pay attention to this step and just tap rapidly on install button.

Emails that users think are safely stored in webmail providers databases are as well parsed, Google for instance is proposing ads for the Gmail users based on their emails content.

Social media is just like a breeding ground for data seekers, Twitter, Facebook, Instagram and many other social platforms are the best way for sharing people's life stories. Users' identities are not the only thing exposed in social media but information with deep level of granularity on people's habits, activities, social behaviour and tendencies are being collected and used with or without the user's will and awareness. WikiLeaks editor-in-chief Julian Assange has branded Facebook an "appalling spying machine", "Here we have the world's most comprehensive database about people, their relationships, their addresses and locations, their communications with each other - all sitting within the United States and all accessible to US intelligence." Says Assange.

As more people get online around the world, especially through their cheap smart phones (of which there are now some 4 billion), data streams will proliferate and most segments of society will reveal themselves through various kinds of social media data. [23] After that all this huge amounts of data is collected about users, it's not of course going to be stored somewhere and remains untouched, it's a gold mine! Imagine how vulnerable and easy to control a person becomes when you own all his personality keys, when you know everything about him and when all his weakness and strengths are exposed. These priceless data is being mined, analyzed and used for different purposes and smart services comes into play:

**Smart e-commerce**

Recommender systems, Behavioral targeting, Customer Profiling, Targeted Marketing and many others are a range of technologies used "against" users for commercial goals, they are techniques invented to investigate individuals' preferences, tendencies, personality, religion, sexual orientation, political views without users' knowledge and take advantage of these information to improve marketing and advertising industry. Some may say there is nothing wrong with bringing better user experience and helping people to easily find what they want! That is a good thing but the bad part is when these massive information collected about individuals and all these "smart" technologies are used to control users' freedom to choose, force them to buy what they don't need, turn them to consumption obsessed and hypnotizing them to pay as much as possible in order to increase firms' sales and revenue.

**Smart search**

Search engines like Google for instance is limiting our search results and personalizing them. If I search for something, and you search for something, even right now at the very same time, we may get very different search results. Even if you're logged out, one engineer told me, there are 57 signals that Google looks at -- everything from what kind of computer you're on to what kind of browser you're using to where you're located -- that it uses to personally tailor your query results. Think about it for a second: there is no standard Google anymore.[24] As Eric Schmidt said, "It will be very hard for people to watch or consume something that has not in some sense been tailored for them." As this filtering algorithms surround us and decide what to show and what to hide from us, the ability to see other point of views is kept away and consequently we are deprived from that natural balance of not only relevant but also a mix of uncomfortable or challenging or important things instead of being captured in a kind of a "bubble" far away from what is actually going in the real world.

**Smart politics**

Political elections' fundamental meaning is the citizen's right to choose his preferred candidate based on the flow of information he gets. So when this citizen is using online services for instance social media, news, search engines that is forcing him to see only one side of the story, then elections seems to become a big joke in our life. "I'm progressive, politically -- big surprise -- but I've always gone out of my way to meet conservatives. I like hearing what they're thinking about; I like seeing what they link to; I like learning a thing or two. And so I was surprised when I noticed one day that the conservatives had disappeared from my Facebook feed. And what it turned out was going on was that Facebook was looking at which links I clicked on, and it was noticing that, actually, I was clicking more on my liberal friends' links than on my conservative friends' links. And without consulting me about it, it had edited them out. They disappeared." Says Eli Pariser [24]

**Smart transportation**

Intelligent transportation systems is one if the services identifying the smart cities, based on using recording devices to collect data including speed, acceleration, braking, seatbelt usage, vehicle status, airbag deployment Hands-free telephone and messaging – Telephone and contact numbers, messages, texts GPS navigation systems – trip data, home site, backtrack data (''breadcrumb'') and other factors. Instrumented transportation systems offer suitable targets for an offender motivated stalking/domestic abuse. First and foremost, the victim/target's privacy is heavily compromised in that access to vehicle systems provides the offender with near complete information on where, when and for how long the victim/target has visited a particular location. It may provide additional information on whom the victim/target called. This privacy violation is a major security risk. Once the motivated offender has a profile and location on the victim/target at all times he or she knows when that victim/target would be most vulnerable to a physical attack.[25]

Eagle eye is a 2008 science fiction movie directed by D.J. Caruso [26], the main idea was about a supercomputer that United States department of defense has engineered named Autonomous Reconnaissance Intelligence Integration Analyst (ARIIA), this intelligent technology was sophisticated enough to auto-collect real time, structured and unstructured big data around the globe and virtually control electronic systems and automated machines. This description is exactly what defines big data analytics systems and smart cities but in the movie the supercomputer became intelligent enough to take its own actions and decide who to kill and who to keep-alive. Are we heading in that direction? Question mark.

## V. CONCLUSION

While big data can provide significant value, it also presents significant risk for our personal security and privacy, whether we like it or not, our personal data is a by-product of our daily lives. Purchases at online and brick-and-mortar retail stores, photos of our license plates taken by surveillance cameras as we drive through intersections, messages posted on social networks—these actions and countless others can tell a story about our lives to those interested in knowing more about us. So something for you to think about: As we adopt these new applications and mobile devices, as we play with these shiny new toys, how much are we trading off convenience for privacy and security? Next time you install something, look at the settings and ask yourself, "Is this information that I want to share? Would someone be able to abuse it?"

What is certain is that what the future will look like depends on the actions we take today. Data is not merely data anymore; it is a commodity that can be bought and sold by corporations, governments, and individuals. [27]

Finally this article was on one hand, a call to the consumers to be aware of how massive data is collected about them and used to tie their liberty and it is up to them to decide the level of data they are comfortable of sharing." Once personal data is out there, it never disappears." Says Adam Tanner in his book [28]. On the other, it is a request to technology companies that are responsible for making our cities smarter to include the respect of personal privacy in their practices and give us the choice!

In sum, the benefits do and will far outweigh the risks when the rights and liberties in a democratic society are observed and protected. The Smart City offers us much. But we must not let it take that which makes us who we are. Difficult and concerted debate on these issues is needed. [25]